\documentclass[a4paper,12pt]{article}
\usepackage{amsfonts}
\usepackage{latexsym}
\usepackage{amsmath}
\usepackage{amssymb}
\usepackage{amssymb}
\usepackage{slashed}
\usepackage{upgreek}
\usepackage{mathrsfs}
\usepackage{bbm}

\usepackage{relsize}
\usepackage{graphicx}

\newcommand{\newsection}{    % Numeration of eqs. is automatic
\setcounter{equation}{0}\section}
\def\appendix#1{\addtocounter{section}{1}\setcounter{equation}{0}
\renewcommand{\thesection}{\Alph{section}}
\section*{Appendix \thesection\protect\indent \parbox[t]{11.15cm}{#1}}
\addcontentsline{toc}{section}{Appendix \thesection\ \ \ #1}}

\font\mybb=msbm10 at 11pt

\def\bb#1{\hbox{\mybb#1}}

\def\bR {\bb{R}}

\newcommand{\bea}{\begin{eqnarray}}
\newcommand{\eea}{\end{eqnarray}}

\newcommand{\be}{\begin{eqnarray}}
\newcommand{\ee}{\end{eqnarray}}

\newcommand{\ba}{\begin{array}}
\newcommand{\ea}{\end{array}}

\begin{document}
\begin{titlepage}
\begin{center}
%\today
\vspace*{-1.0cm}
%\hfill hep-th/yymmnnn \\
%\hfill UB-ECM-PF-06-43 \\
\hfill DMUS--MP--16/15 \\

\vspace{2.0cm} {\Large \bf On supersymmetric  Anti-de-Sitter, de-Sitter and Minkowski flux backgrounds} \\[.2cm]

\vskip 2cm
U. Gran$^1$,   J. B.  Gutowski$^2$ and G. Papadopoulos$^3$
\\
\vskip .6cm

\begin{small}
$^1$\textit{Department of Physics, Chalmers University of Technology\\
SE-412 96 G\"oteborg, Sweden\\
Email:ulf.gran@chalmers.se}
\end{small}\\*[.6cm]

\begin{small}
$^2$\textit{Department of Mathematics,
University of Surrey \\
Guildford, GU2 7XH, UK \\
Email: j.gutowski@surrey.ac.uk}
\end{small}\\*[.6cm]

\begin{small}
$^3$\textit{  Department of Mathematics, King's College London
\\
Strand, London WC2R 2LS, UK.\\
E-mail: george.papadopoulos@kcl.ac.uk}
\end{small}\\*[.6cm]

\end{center}

\vskip 3.5 cm
\begin{abstract}

We test the robustness of the conditions required for the existence of (supersymmetric) warped flux anti-de Sitter,  de Sitter, and Minkowski backgrounds in supergravity theories
 using as  examples suitable foliations of anti-de Sitter spaces.
We find that
 there are supersymmetric de Sitter solutions in supergravity theories including
 maximally supersymmetric ones  in 10- and 11-dimensional supergravities.
Moreover, warped flux Minkowski backgrounds can admit Killing spinors which are not Killing on the Minkowski subspace
and therefore cannot be put in a factorized form.

\end{abstract}

\end{titlepage}

%%%%%%%%%%%%%%%%%%%%%%%%%%%%%%%%%%%%%%%%%%%%%%%%%%%%%%%%%%%%%%%%%%%%%%%%%%

%%%%%%%%%%%%%%%%%%%%%%%%%%%%%%%%%%%%%%%%%%%%%%%%%%%%%%%%%%%%%%%%%%%%%%%%%%
\section{Introduction}

 Anti-de Sitter (AdS),  de Sitter (dS) and Minkowski (Mink) warped flux backgrounds have an important role in a variety of applications in the context of supergravity,
string theory, M-theory and gravity/CFT correspondences, for reviews see e.g.~\cite{duff, grana, maldacena, strominger3,  klemm}. Because of this, there has been much progress
towards the
classification of such backgrounds with special focus on  AdS solutions, see e.g.~\cite{dario1}-\cite{passias1}. A different classification method for AdS and Mink backgrounds has also been
proposed in \cite{mads, iibads, iiaads} which  makes no assumptions on either the form of the fields or that of the Killing spinors apart from imposing on the former the symmetries of AdS or Mink subspaces.

During the historical development of the area, and in efforts to find  explicit solutions, some assumptions have been
 made on  the properties of such backgrounds and their Killing spinors in addition to the usual requirement that the fields must be invariant under the isometries of the AdS, dS or  Mink subspaces.
 One of them has been the requirement that the spacetime Killing spinors factorize into Killing spinors of the AdS, dS and Mink subspaces and to Killing spinors on the
 internal space. These additional assumptions have consequences and a priori put additional restrictions on the type of solutions that can occur as well as the counting of their supersymmetries.

The purpose of this paper is to provide a concise example based description of issues that arise in the understanding of the properties of (supersymmetric) warped flux AdS,  dS and Mink backgrounds especially used in, but not exclusively
related to, the classification of all such solutions.
These issues include the exploration of questions regarding the existence of such backgrounds in supergravity theories,  the number of supersymmetries that  they preserve, and the
way that these supersymmetries can be counted.  The last two questions are closely related to the question of whether all Killing spinors can be factorized
into Killing spinors on the AdS,  dS and Mink subspaces  and Killing spinors on the internal spaces\footnote{We adapt the terminology of compactifications where
AdS,  dS or  Mink is the lower-dimensional spacetime and the transverse space is the internal space.  However no restrictions are put on the internal
space like the requirement to be compact.}. Such questions have originally been raised in \cite{mads, iibads, iiaads} for warped  flux  AdS and Mink backgrounds. Here the examples will further elucidate the issues involved and the dS backgrounds will also be included in the analysis.

 Our analysis leads to some unexpected  conclusions regarding the existence of supersymmetric dS, AdS  and Mink backgrounds\footnote{Note that the spaces $AdS_2$ and $dS_2$ are isometric. Their metrics differ by an overall sign which is not an observable.}. There are supersymmetric dS solutions in supergravity theories. In particular, for every supersymmetric AdS$_k$, $k>2$, warped flux solution
 of a supergravity theory there is at least one {\it supersymmetric  warped flux dS$_n$ solution}, $2\leq n<k$, preserving the same number of supersymmetries.  For example, there are maximally supersymmetric $dS_n$ solutions\footnote{All these solutions are locally isometric to $AdS_7\times S^4$ in agreement with the classification results of maximally supersymmetric solutions in \cite{maxsusy}.} in 11-dimensional supergravity for every $2\leq n<7$, and similarly in IIB there are maximally supersymmetric $dS_n$ solutions for every $2\leq n<5$.

 One can also show that similar conclusions to those
described above for dS solutions also hold for warped flux Mink backgrounds. In addition, one can show explicitly that some supersymmetric
  warped flux Mink solutions admit Killing spinors which {\it depend non-trivially on the coordinates of the Mink subspace}. Such Killing spinors are not invariant under all of the symmetries
 of Minkowski space but nevertheless solve the KSEs of the supergravity theories. Clearly such Killing spinors cannot be written in a factorized form as in (\ref{fact}) and therefore
 their exclusion leads to an incorrect counting of supersymmetries.

 A similar analysis can be done for warped flux AdS solutions to conclude that for every supersymmetric AdS$_k$ background, there are warped flux AdS$_n$ backgrounds with $n<k$ that
 preserve the same number of supersymmetries.  So in 11-dimensional supergravity there are maximally supersymmetric warped flux AdS$_n$ solutions for $n\leq 7$ and in IIB for $n\leq 5$.

 Before we proceed to explain how we obtain these results let us state our assumptions.  It is known for some time that there are no smooth warped flux Mink and dS solutions of 10- and 11-dimensional
 supergravity theories with compact without boundary internal space \cite{gibbons1, maldacena1}, see also section 3. So our results on Mink and dS solutions are local and apply to a suitable open set or when the internal spaces
 can be taken  to be non-compact. On the other hand there are smooth warped flux AdS compactifications and therefore our results can apply in general in this case.

The difference between our results and the prevailing line of thought in some of the literature regarding the properties of supersymmetric warped flux AdS, dS and Mink backgrounds
can be explained by the way that the supergravity KSEs are solved in the two cases. Typically in the literature, the KSEs are solved after imposing a factorization of the spacetime Killing
spinors in terms of suitable Killing spinors\footnote{These are postulated to solve a Killing spinor equation of the type described in (\ref{lkse}).} on the  AdS, dS or Mink subspace and  some Killing spinors on the internal space. In our approach, which has been described in detail
for AdS and Mink backgrounds in \cite{mads, iibads, iiaads}, no such assumption is made.  Instead, the spacetime KSEs are integrated over the AdS or Mink subspaces to find the explicit dependence on  the coordinates of these spaces,  and then the remaining KSEs on the internal space are identified. The Killing spinors do not generically factorize as in (\ref{fact}) and (\ref{lkse}).
Presumably a similar analysis can be done for warped flux dS backgrounds. Our results explicitly show that the problem of non-existence of real Killing spinors
on the dS space, required by the factorization of Killing spinors,  is not a barrier for the existence of Killing spinors over the whole spacetime.

This paper is organized as follows. In section 2, we present a class of foliations of AdS spaces with leafs either AdS, dS or Mink subspaces and prove that
 the metric of the original AdS space takes a suitable warped form. In section 3, we review a non-existence theorem for dS and Mink backgrounds in supergravity theories. In section 4,
 we investigate the existence of supersymmetric warped flux AdS, dS and Mink backgrounds in supergravity theories, and in section 5 we give our conclusions.

\newsection{Foliating and Warping AdS }

The examples we shall present to describe the properties of warped flux AdS, dS and Mink backgrounds utilize the property that the AdS spaces can be foliated
with AdS, dS and Mink subspaces.  This has been explored before in the literature \cite{strominger}-\cite{slicex}.  Here we shall give a description which is suitable for
our purposes and resembles the form used in \cite{ajg}.  It is known that AdS$_{n+1}$ space can be described as a conic
\bea
-t_1^2-t_2^2+ x_1^2+ \dots + x_n^2=-\ell^2~,
\label{con}
\eea
where $\ell$ is the radius. Using polar coordinates for $(t_1, t_2)$ and $(x_1, \dots, x_n)$, separately, one can describe the standard metric on AdS$_{n+1}$
 in global coordinates.

Furthermore AdS$_{n+1}$ can be seen as a foliation with leafs given by AdS$_n$.  To see this, first set
\bea
t_1=\rho \cos(t)~,~~~t_2=\rho \sin(t)~,~~~\rho\geq 0
\eea
and thus the equation for the conic becomes
\bea
-\rho^2+x_1^2+ \dots+ x_n^2=-\ell^2~,
\eea
and the induced metric on AdS$_{n+1}$ reads
\bea
ds^2=-\rho^2 dt^2-d\rho^2+ dx_1^2+\dots+ dx_n^2~,
\eea
where $t$ is not taken  to be periodic.
Next set
\bea
\rho = \ell r \cosh y , \qquad x_n=\ell \sinh y \ ,
\eea
for $r \geq 0$, and
\bea
\qquad x_j = \ell w_j \cosh y \qquad  {\rm for} \ {j=1, \dots, n-1} \ .
\eea
In these co-ordinates, the equation for the conic becomes
\bea
-r^2+ w_1^2+ \dots + w_{n-1}^2=-1 \ .
\eea
The resulting metric is
\bea
ds^2(AdS_{n+1})=\ell^2 dy^2+ \ell^2 \cosh^2y \big(-r^2 dt^2-dr^2+dw_1^2+ \dots + dw_{n-1}^2 \big) 
\eea
and hence
\bea
ds^2(AdS_{n+1})=\ell^2 dy^2+ \ell^2 \cosh^2y\, ds^2(AdS_n)~,~~~y\in \bR~,
\label{adswarp}
\eea
which is clearly in a warped product form with warp factor $\cosh^2y$.
Observe that in this choice of parametrization the metric of both
AdS$_{n+1}$ and AdS$_n$ are written in global coordinates.  Therefore the whole of AdS$_{n+1}$ is foliated with  AdS$_n$ subspaces.  Clearly the
procedure can be repeated and therefore an AdS space can be foliated with any AdS subspace of lower dimension.

Next let us consider the foliation of AdS$_{n+1}$ with dS$_n$. To solve the conic equation  (\ref{con}),  introduce polar coordinates for
$(x_1, \dots, x_n)$, which yields
\bea
-t_1^2-t_2^2+ r^2=-\ell^2~,
\eea
where $r^2= x_1^2+ \dots+ x_n^2$.  The induced metric then becomes
\bea
ds^2(AdS_{n+1})=-dt_1^2-dt_2^2+dr^2+ r^2 ds^2(S^{n-1})~.
\eea
Next set $t_2=\rho \sinh \xi$ and $r=\rho \cosh \xi$, with $\rho\geq 0$. The equation for the conic reads
\bea
-t_1^2+\rho^2=-\ell^2~,
\eea
and the induced metric is
\bea
ds^2(AdS_{n+1})=-dt_1^2+ d\rho^2- \rho^2 d\xi^2+ \rho^2 \cosh^2\xi\, ds^2(S^{n-1})~.
\eea
Finally setting $\rho=\ell \sinh y$ and $t_1=\ell \cosh y$, the metric can be rewritten as
\bea
ds^2(AdS_{n+1})=\ell^2 dy^2+ \ell^2 \sinh^2 y \, ds^2 (dS_n)~,
\label{dswarp}
\eea
where
\bea
ds^2 (dS_n)=-d\xi^2+\cosh^2\xi\,  ds^2 (S^{n-1})~
\eea
is the metric on dS written in global coordinates.  Clearly, the metric of  AdS$_{n+1}$  is in warped form. Since it is required that $\rho\geq 0$, we have to make the restriction
that $\sinh y\geq 0$.  Furthermore notice that $t_1\geq 0$,  and as a result only part of AdS$_{n+1}$ is foliated with dS$_n$ subspaces.  Alternating
AdS and dS foliations of an AdS space, one concludes that (a suitable part of) AdS spaces can be foliated with dS spaces
of any lower dimension.

To construct  a foliation of  AdS$_{n+1}$  with Minkowski subspaces, we can consider the Poincar\'e patch.  In such a case, the metric of  AdS$_{n+1}$
can be written as
\bea
ds^2(AdS_{n+1})= \ell^2 dy^2+ \ell^2 e^{2y} ds^2(Mink_n)~,
\label{fwarp}
\eea
which is of course in warped form. Only part of AdS$_{n+1}$ is  foliated in this way.

\newsection{ Existence of dS and Minkowski solutions}

\subsection{Revisiting  a non-existence theorem}

It is known that there are restrictions on the existence of smooth warped Minkowski and dS solutions with fluxes in 10- and 11-dimensional supergravities \cite{gibbons1, maldacena1}.
Here, we review the argument from the point of view of the Hopf maximum principle.   To be concrete, consider the metric of such a background
\bea
ds^2= e^{2\Phi} ds^2(M^n)+ ds^2(N^{D-n})~,
\eea
where $M^n$ stands for $\bR^{n-1,1}$, $AdS_n$ or $dS_n$ and $N^{D-n}$ is the internal space and $\Phi$ is a function of $N^{D-n}$ only.  The latter is required for the
background to preserve the isometries of $M^n$.  Imposing the symmetries of $M^n$ onto the full solution, one can find expressions for the remaining form field strengths
of all (massive) IIA, IIB and 11-dimensional supergravity theories.

Next, suppose we are considering only supergravity theories without higher order curvature corrections and sources.  In such a case,   the Einstein equation in all these theories reveals that the warp factor satisfies the equation
\bea
\nabla^2 e^{n \Phi} = q e^{(n-2)\Phi}  R(M^n)+ e^{n \Phi} S(F)~,
\label{keyeqn}
\eea
where $R(M^n)$ is the Ricci scalar of $M^n$, $S(F)\geq 0$ is a semi-definite function of the metric and fluxes $F$, and $P$  depends on the fields and $q$ is a constant $q>0$.

Now if $R(M^n)\geq 0$, which is the case for Minkowski space and dS, the the right-hand side of the above equation is positive
semi-definite.  In these two cases, one can apply the Hopf maximum principle. Assuming that all the data are smooth, including the warp factor $\Phi$,
and that the internal space is compact without boundary,  one finds that the only solution is  that $\Phi$ is constant. Consistency then requires that
$R(M^n)=0$ and $S(F)=0$.  Therefore the equation has smooth solutions if and only if   $M^n=\bR^{n-1,1}$ and the fluxes are restricted by $S=0$.
The latter condition for 10- and 11-dimensional supergravities requires that the fluxes vanish.  Therefore, one concludes the following \cite{gibbons1, maldacena1}
\begin{itemize}

\item There are no smooth warped flux dS solutions of 10- and 11-dimensional supergravities with compact without boundary internal space.

\item There are no smooth  warped flux Mink solutions to 10- and 11-dimensional supergravities with non-trivial fluxes and
compact without boundary internal space.

\end{itemize}
We remark that this non-existence theorem is a consequence of the field equations and therefore it is valid also for non-supersymmetric solutions.
In the case of AdS spaces, the above argument does not apply as the right-hand-side of (\ref{keyeqn}) is not positive semi-definite.
This is in agreement with the existence of many smooth AdS solutions with compact  without boundary internal space.

The Hopf maximum principle can be applied also to spaces with boundary but it requires some additional assumptions. To see this, suppose
that $N^{D-n}$ is compact but with a smooth boundary. It turns out that (\ref{keyeqn}) again does not admit solutions provided that $\Phi$
has a maximum in the interior. This is because if such a maximum existed as a critical point the left-hand-side would have been strictly
negative while the right-hand-side is positive semi-definite. The implications for Minkowski and de Sitter backgrounds are as those stated
above. However in this case, there may be solutions which do not exhibit a maximum\footnote{Other critical points are allowed, like e.g.~saddle points.} in the interior. In such a case, the maximum will occur
at the boundary of the internal space. 

\subsection{Robustness of the assumptions}

 To test the assumptions made for the proof of the non-existence theorem in the previous section,  first notice that there exist     smooth  dS and Mink solutions in supergravity theories with  non-compact internal spaces. Examples include the planar   M-branes and D3 brane solutions and for dS solutions see  \cite{gibbons3}. The foliations of AdS space described in section 2 provide many more examples. In what
follows we shall examine more closely the robustness of the assumptions made in the proof of the non-existence theorem in the previous section using the foliations of AdS space.

  Clearly,  higher order corrections to supergravity coming from string theory or some  other modification of supergravity theory that includes higher
curvature terms  can change the conclusions as they may interfere with the positive semi-definiteness of the right-hand-side of (\ref{keyeqn}).
A similar conclusion can be reached by adding additional source terms into the Einstein equations or by removing the smoothness
assumptions on the fields.

Therefore, for the sake of argument, let us stay within supergravity theory and insist on the smoothness of the fields.  In the case of Mink solutions, one can consider one of the smooth AdS solutions of 10- and 11-dimensional supergravity theories, like the $AdS_5\times S^5$ solution in IIB or the
$AdS_7\times S^4$ solution of 11-dimensional supergravity. Then write the $AdS$ subspace as a warped product of Minkowski space as described in (\ref{fwarp}) in section 2.  The warp factor $e^{2y}$ is smooth.
Next if the $y$ coordinate is extended over the whole real line, the internal space $N$ is not compact.  On the other hand if it is restricted to an interval it attains a maximum value at the boundary as expected from the Hopf maximum principle. If one tries to periodically identify the $y$ direction so that the internal space is compact,
then the warp factor is not continuous.  In all cases, the solutions behave as required by the general analysis.

Next consider the dS case. Again let us begin with a smooth AdS solution in 10 or 11 dimensions and then warp the $AdS$ with a $dS$ subspace as described in (\ref{dswarp}) in section 2.
In this case, the warp factor is $\sinh^2y$ with $y>0$.  The warp factor vanishes at $y=0$, but this is a coordinate singularity as AdS is non-singular. The remarks made
for the Mink solutions also apply to this case. The solutions behave as expected from the general theory. Moreover none of the assumptions
for the validity of the non-existence theorem can be weakened, as otherwise there are solutions.

\newsection{Supersymmetric AdS, dS and Mink solutions }

\subsection{Two approaches to supersymmetry}

\subsubsection{Factorization}

There are two approaches to  investigate the supersymmetric warped flux   AdS, dS  and Mink backgrounds of a supergravity theory.
The traditional point of view is to consider these backgrounds as compactifications to a lower dimensional theory and
{\it assume} that the spacetime Killing spinors factorize
\bea
\epsilon=\sum_r \eta_r\otimes \xi_r~,
\label{fact}
\eea
where  $\eta_r$ are solutions of the KSE\footnote{In fact, the KSE takes the form $\tilde\nabla_\mu \eta=\lambda \gamma_\mu\otimes K\, \eta$, where $K$ is a matrix
that arises from the factorization of spacetime gamma matrices to those on AdS, dS or Mink and those of the internal space. But this has no effect in the arguments that follow.}
\bea
\tilde\nabla_\mu \eta=\lambda \gamma_\mu \eta~,
\label{lkse}
\eea
on AdS, dS or Mink with Levi-Civita connection $\tilde\nabla$ and $\xi_r$ are solutions of the KSEs on the internal space,
where $R(M^n)=-\lambda^2$ and $\gamma$ are the scalar curvature and  gamma matrices on AdS, dS or Mink spaces, respectively. The  KSEs of the internal space are derived after enforcing (\ref{fact}) on the supergravity KSEs and  using (\ref{lkse}).  Several different factorizations have appeared in the literature depending on the number of terms in the sum in (\ref{fact}) and
the reality properties of the spinors.  All of them lead to constant Killing spinors along the Mink subspaces, in the standard coordinates and frame,  of the associated warped flux backgrounds.
We shall refer to this as the ``factorization'' approach or method.

\subsubsection{Integration}

There is another method to solve the  KSEs of supergravity theories without making any assumptions on the form of the Killing spinors,  on the internal space $N$  or on the fields, apart from imposing on the latter the isometries of the AdS, dS or Mink subspaces  \cite{mads, iibads, iiaads}.  For this, one substitutes the fields of the associated backgrounds into the supergravity  KSEs
and  solves them along the AdS, dS  or  Mink subspaces. Then one collects the remaining KSEs along the internal space.
This investigation has been carried out for AdS and Mink solutions  where the a priori number of supersymmetries
preserved by such backgrounds were counted. The novelty of this approach is that it is completely general
and the only assumption that is made is the invariance of the fields under the isometries of AdS, dS or Mink.   We shall refer to this
as the ``integration'' approach or method.

As we shall explain, with explicit examples below, the factorization approach contains hidden assumptions.  As a result, one does not get
the most general form of the Killing spinor in this way.  Furthermore, it has consequences for the existence of supersymmetric dS backgrounds.

\subsection{Minkowski}

One of the apparent consequences of the application of the factorization method to Mink solutions is that all the Killing spinors
of these solutions do not depend on the standard coordinates of the Mink space. This is because (\ref{lkse}) in the standard coordinates and frame becomes $\partial\eta=0$ and so all solutions $\eta$
are constant. Thus the only expected dependence of the spacetime Killing spinors is on the coordinates of the internal space.

Next consider  the integration approach to solving the KSEs of supergravity theories that has been investigated in detail  for Mink backgrounds
in \cite{mads, iibads, iiaads}. There it was found that apart from the constant Killing spinors  there may exist some additional ``mysterious'' ones which depend
non-trivially on the standard coordinates of Mink.  In particular, these Killing spinors are in the kernel of a Clifford algebra operator $\Xi^2$ but not in the kernel of $\Xi$,
 which is constructed from the metric and fluxes, see \cite{mads} for the definition of $\Xi$ in 11 dimensions. Naturally, this observation has led the authors of these papers to argue that in general
the Killing spinors  of warped flux Minkowski backgrounds do not factorize as in (\ref{fact}) and (\ref{lkse}).

It turns out that there are examples that clarify and confirm the presence of these additional mysterious Killing spinors in warped  flux Mink solutions.
These\footnote{These examples emerged during a discussion of one of us, GP,  with A.~Tomasiello. }    are the maximally supersymmetric AdS$_{n+1}$ backgrounds of IIB and 11-dimensional supergravity now viewed in the Poincare patch as warped flux Mink$_n$ solutions,
see also section 2. Such solutions in Poincare coordinates, and in an appropriate frame,  have apart from constant Killing spinors also Killing spinors that depend non-trivially
of the coordinates of the Mink$_n$ subspace.  The latter are necessary for the correct counting of supersymmetries. 

It would be of interest to find whether there are other warped flux  Mink solutions of supergravity theories
which are not foliations of AdS spaces and admit Killing spinors which depend non-trivially on the standard coordinates of the Mink space. This question can be
answered on a case by case basis since given the fluxes and the metric, the kernels of the  Clifford algebra operators $\Xi^2$ and $\Xi$  can be computed explicitly.

\subsection{AdS}

The warped AdS backgrounds of 10- and 11-dimensional supergravity theories have been investigated in complete generality in \cite{mads, iibads, iiaads}.
The question of the factorization of Killing spinors as in (\ref{fact}) has also been investigated where it was found that in general they
do not factorize. This is also supported by the following indirect argument.  It is well known that at very large AdS radius the warped flux AdS backgrounds become Mink
backgrounds.  This limit is smooth on both the fields and the supergravity KSEs. Taking the limit in any factorization method, the only solutions of (\ref{lkse}) are constant
spinors in the standard coordinates of the Mink subspace in the limit. However, as we  have explained above there are counter-examples to  this.

Here we would like to point out an application of the foliation of AdS spaces with AdS subspaces. As the maximally supersymmetric,
and other, AdS backgrounds of supergravity theories can be foliated with AdS subspaces, one concludes that 11-dimensional supergravity admits
maximally supersymmetric\footnote{Note that in the non-existence theorem of \cite{passias2} for AdS$_6$ solutions,
the additional assumption of compactness of the internal space was made.
This example illustrates that such an additional assumption is necessary to rule out these solutions.} $AdS_k$ solutions
for every $2\leq k\leq 7$ and similarly the IIB supergravity admits maximally supersymmetric $AdS_k$ solutions
for every $2\leq k\leq 5$.
Of course all these solutions are locally isometric to either $AdS_4\times S^7$ or $AdS_7\times S^4$ in 11 dimensions or
$AdS_5\times S^5$ in IIB, all of which are maximally supersymmetric solutions in agreement with the classifications results of \cite{maxsusy}.

\subsection{dS}

The KSE (\ref{lkse}) does not admit any real solutions on the dS spaces\footnote{To resolve this problem, in \cite{anous} it was suggested to use a conformal KSE instead of (\ref{lkse}). However
as we shall see this is not necessary in supergravity.}.  As a result based on the  factorization
method, it is often stated in the literature   that there do not exist supersymmetric warped flux dS backgrounds in supergravity theories with a real supersymmetry parameter.
This would exclude for example the existence of supersymmetric dS solutions  in  11-dimensional supergravity.

The integration method, unlike the cases for AdS and Mink solutions,  has not been applied to warped flux  dS backgrounds.  Nevertheless,
it is possible via explicit examples to argue that there are such supersymmetric dS backgrounds.
Indeed consider again supersymmetric AdS backgrounds of 10- and 11-dimensional supergravity theories for which the AdS spaces are foliated with
a dS subspace as in (\ref{dswarp}). The resulting metric is a warped product of dS$_n$ space with a transverse space which includes the $y$ coordinate,
\bea
ds^2=\ell^2 e^{2\Phi(z)}\, \sinh^2 y \,\,ds^2(dS_n)+\ell^2 e^{2\Phi(z)}\, dy^2+ ds^2(N^{D-n-1})~.
\eea
However,  since this is actually an  AdS$_{n+1}$ solution, it can preserve some supersymmetry.  In fact, if we foliate the maximally supersymmetric $AdS_7\times S^4$ solutions, we
conclude that there are maximally supersymmetric warped flux dS$_n$ backgrounds in 11-dimensional supergravity theory for $2\leq n<7$.
For this we can appropriately alternate AdS and dS foliations of AdS$_7$.
A similar conclusion can be reached in IIB based on the maximally supersymmetric $AdS_5\times S^5$ solution for $2\leq n<5$. Of course for such backgrounds to be viewed as dS solutions
the internal spaces are not compact as expected from the general non-existence theorem in section 3.

Further dS solutions can be constructed in 10- and 11-dimensions using  chains of dualities.  For example, once can consider $AdS_7\times S^4$, view $S^4$ as a warped product foliation
of $S^3$ and then reduced on the Hopf fibre of $S^3$ to find a new solutions in IIA.  This can be re-interpreted as $dS_n$ solution for $n<7$.

\section{Concluding remarks}

The foliations of AdS spaces with AdS, dS and Minkowski subspaces can be used as a laboratory to test the various assumptions
made on the construction and properties of warped flux AdS, dS and Minkowski  backgrounds  in supergravity theories.
Although these examples are of a particular type, nevertheless they provide answers to a number of subtle questions regarding the properties
of these backgrounds. In particular, we have found that warped flux Minkowski solutions admit Killing spinors which can depend non-trivially on the standard coordinates
of Minkowski space and that there are supersymmetric warped flux dS solutions in supergravity theories, some of which preserve maximal
supersymmetry.

These results have a profound application for the classification of all warped flux compactifications as they question one of the
basic assumptions made for their construction; that of the factorization of the Killing spinors in (\ref{fact}). As a result of this,
the generality of the factorization approach as applied to the classification of such backgrounds must be re-examined. In any case,
for AdS and Mink solutions there is the alternative of the integration method for all cases.

The supersymmetric warped flux dS solutions are of particular interest. This is because of cosmological considerations in the context of string and M-theory and the dS/CFT correspondence  \cite{strominger2}. It is not apparent that the analysis we have done is sufficient
to address the question of existence of appropriate warped flux dS backgrounds suitable for such applications. The main issues of incorporating the dS space in the context of string theory and M-theory, either in a cosmological setting or in dS/CFT, are associated with the non-existence theorems
 of \cite{gibbons1, maldacena1} before the question of supersymmetry arises.  Because of this  constructions such as that in \cite{kallosh} may be necessary. However,
 what our results show using the foliations of AdS spaces is that warped flux dS backgrounds can be supersymmetric in a similar way to that of AdS or Mink backgrounds.
 In fact, the non-existence of real solutions to (\ref{lkse}) can be interpreted as breaking all  supersymmetry at the compactification scale instead of prohibiting
 the existence of supersymmetric solutions in the higher dimensional theory.
   So it would be of interest to integrate the KSEs of supergravity theories on
warped flux dS backgrounds and thus extend the integration method to solving the KSEs in this case so that the number of supersymmetries preserved by such backgrounds
can be counted and a more systematic search for solutions can be initiated.

\vskip 0.5cm
%\newpage
\noindent{\bf Acknowledgements} \vskip 0.1cm
\noindent   We would like to thank A.~Tomasiello for very helpful discussions and D.~Martelli for comments.
UG is supported by the Swedish Research Council. JG is supported by the STFC grant, ST/1004874/1. GP is partially supported by the  STFC rolling grant ST/J002798/1.
JG would like to thank the Department of Mathematical Sciences, University of Liverpool for hospitality during which part of this work was completed.
\vskip 0.5cm

\vskip 0.5cm
%\newpage
\noindent{\bf Data Management} \vskip 0.1cm

\noindent No additional research data beyond the data presented and cited in this work are
needed to validate the research  findings in this work.

\vskip 0.5cm

\setcounter{section}{0}\setcounter{equation}{0}

\end{document}